
\def\s#1{{\textstyle{#1}}}

\def\A{{\cal A}}
\def\B{{\cal B}}
\def\G{{\cal G}}
\def\L{{\cal L}}
\def\BB{\overline{\cal B}}
\def\JJ{\overline J}
\def\xx{\overline x}
\def\diag{\hbox{diag}\,}
\def\bra{\langle}
\def\ket{\rangle}

\magnification=1200
\rm
\tolerance=1000
\pretolerance = 5000
\centerline{\bf EXTENDED DE SITTER THEORY OF}
\medskip
\centerline{\bf TWO-DIMENSIONAL GRAVITATIONAL FORCES}
\bigskip
\bigskip
\centerline{ Daniel Cangemi $^{\ddagger}$ and Gerald Dunne $^{\dagger}$}
\bigskip
\bigskip
\centerline{\it $^{\ddagger}$Center for Theoretical Physics, Laboratory for
Nuclear Science,}
\centerline{\it and Department of Physics}
\centerline{\it Massachusetts Institute of Technology}
\centerline{\it Cambridge, MA 02139}
\bigskip
\bigskip
\centerline{\it $^{\dagger}$ Department of Physics}
\centerline{\it University of Connecticut}
\centerline{\it Storrs, CT 06269}
\bigskip
\bigskip
\bigskip
\bigskip
\bigskip
\bigskip
\centerline {\bf ABSTRACT}
\bigskip
\bigskip
\bigskip
\bigskip
We present a simple unifying gauge theoretical formulation of gravitational
theories in two dimensional spacetime. This formulation includes the effects
of a novel matter-gravity coupling which leads to an extended de Sitter
symmetry algebra on which the gauge theory is based. Contractions of this
theory encompass previously studied cases.
\vfill
\noindent CTP\#2228 \par
\noindent UCONN-93-6 \hfill August 1993 \par
\noindent hep-th/9308021
\eject
\noindent{\bf I. INTRODUCTION}
\bigskip
Recent work has shown that gravitational theories in $1+1$ dimensions may
be consistently formulated as gauge theories~[1]. For string-like models of
``dilaton" gravity~[2] one uses the Poincar\'e group $ISO(1,1)$~[3] or
alternatively the extended Poincar\'e group $\widehat{ISO}(1,1)$~[4,5],
while the constant curvature models involve the de Sitter group
$SO(2,1)$~[6].
Such gauge models are of interest because of the deep connection between
diffeomorphism invariance and gauge invariance, and because they provide a
picture of two dimensional gravity complementary to the more geometrical
string-like approach~[7]. Moreover, there is some hope that ``black hole" type
solutions may yield important new information about the quantum mechanics of
realistic four dimensional black holes~[8]. It has also been shown that these
gauge theories arise as dimensional reductions of $2+1$ dimensional gauge
theories of gravity~[9].

In this paper we seek to unify these gauge theoretical models by considering
the effects of a novel type of matter-gravity
interaction in the constant curvature de Sitter model. This interaction
involves matter being minimally coupled to a gauge field whose field
strength two form is everywhere proportional to the area two form. While the
physics is very different, at least formally such an interaction is reminiscent
of the classic Landau problem of dynamics in the presence of a uniform magnetic
field, an interaction which has an interesting generalization to nonzero
constant curvature surfaces~[10], and even to higher genus surfaces~[11]. This
type of matter-gravity coupling has already been considered
in the zero curvature theory~[5]. Here we generalize to nonzero constant
curvature and show that this serves as a unifying model from which all
other previously studied gauge theories of $1+1$ dimensional gravity may be
obtained by contraction. In particular, this model clarifies the significance
of the extension needed in the Poincar\'e theory.

This paper is organized as follows. In the remainder of the Introduction we
state our conventions for the geometric quantities to be used throughout this
paper. In Section {\bf II} we discuss the matter-gravity coupling in both
Lagrangian and Hamiltonian approaches, showing how the extended de Sitter
algebra arises as a symmetry algebra in each case. Section {\bf III} is
devoted to a solution of the equations of motion by direct integration, making
use of the de Sitter symmetries. In Section {\bf IV} we show how the theory may
be reformulated in an elegant and compact form as a gauge theory based on the
extended de Sitter group. The gauge formulation of the matter-gravity
coupling is analyzed in Section {\bf V}, and we conclude with some brief
comments in Section {\bf VI}.

We shall use the convention that lower case Greek and Latin indices
refer respectively to space-time and tangent space components.
Space-time indices are raised and lowered with the metric tensor
$g_{\mu \nu}$, and tangent space indices with the Minkowskian metric
$h_{a b}=\diag(1,-1)$. The {\it Zweibein} $e^a_\mu$ is related to the
metric tensor $g_{\mu \nu}$ by
$$
g_{\mu \nu}~=~e^a_\mu ~e^b_\nu ~h_{a b}
\eqno(1.1)$$
The spin-connection, when defined in terms of the {\it Zweibein}, is
given by
$$
\omega_\mu= {1\over \sqrt{-g}}\epsilon^{\alpha \beta} e^a _\mu h_{a
b} \partial_\alpha e^b _\beta
\eqno(1.2)$$
Here $\epsilon^{\mu \nu}$ is the {\it numerical} anti-symmetric
tensor, with sign convention $\epsilon^{0 1} =1$. Thus $\epsilon^{\mu \nu}
/\sqrt{-g}$ is a contravariant tensor, where
$$\eqalignno{
g&= \det g_{\mu \nu} \cr
\sqrt{-g}&= \det e^a_\mu &(1.3)\cr
&=-{1\over 2} e^a_\mu e^b_\nu \epsilon^{\mu \nu} \epsilon_{a
b}\cr}
$$
The scalar curvature $R$
is related to the spin connection by
$$
{1\over\sqrt{-g}} \epsilon^{\mu\nu} \partial_\mu \omega_\nu = {1\over2} R
\eqno(1.4)$$
In conformally flat coordinates the space-time metric is
$$
g_{\mu \nu}~=~e^{-2 \sigma} ~h _{\mu \nu}
\eqno(1.5)$$
and the corresponding {\it Zweibein} and spin connection are
$$\eqalignno{
e^a _\mu ~&=~e^{-\sigma} \delta ^a _\mu \cr
\omega_\mu&=\epsilon_\mu{}^\nu \partial_\nu \sigma&(1.6)\cr}
$$
In these coordinates, the relation (1.4) reduces to an equation relating
the curvature $R$ and the conformal factor $\sigma$ :
$$
\partial_+ \partial_- \sigma = {1\over4} R e^{-2 \sigma}
\eqno(1.7)$$
Here $\partial_+ \equiv \partial /\partial x^+$, $\partial_- \equiv
\partial /\partial x^-$, where $x^\pm$ are light-cone coordinates
$x^{\pm} = {1\over\sqrt{2}} (x^0 \pm x^1)$. Later, we shall be interested
 in the coupling of matter to gravitational fields with {\it constant}
curvature, $R = \Lambda$, in which case equation (1.7) is just the familiar
Liouville equation. The general solution is
$$
\sigma = \log {1 + \s{\Lambda\over4} F(x^+) G(x^-) \over F'(x^+) G'(x^-) }
\eqno(1.8)$$
for any functions F and G. We can, however, always transform it to the
solution
$$
\sigma = \log (1 + \s{\Lambda\over4} x^+ x^-).
\eqno(1.9)$$
\bigskip
\noindent {\bf II. ``MAGNETIC" MATTER-GRAVITY INTERACTION}
\bigskip
\noindent {\bf II.1 Lagrangian Formalism}
\medskip
The conventional action for a point particle of mass $m$, moving on the
world line $x^\mu (\tau)$, is proportional to the arc length
$$
I_m = - m \int d\tau \sqrt{\dot x^\mu(\tau) g_{\mu\nu}(x(\tau)) \dot
 x^\nu(\tau)}
\eqno(2.1)$$
Here, and throughout this paper, the overdot denotes differentiation with
respect to the world line parameter $\tau$. It is, however, possible to
consider an additional parametrization invariant term in the action~[5].
This term is linear in $\dot x^\mu$,
$$
I_{\rm coupling} =- \int d\tau \dot x^\mu(\tau) \Bigl( \A
\omega_\mu(x(\tau))
 + \B a_\mu(x(\tau)) \Bigr)
\eqno(2.2)$$
where $\A$ and $\B$ are constants, $\omega_\mu$ is the spin-connection
and
 $a_\mu$ a one-form
defined by
$$
{1\over\sqrt{-g}} \epsilon^{\mu\nu} \partial_\mu a_\nu = 1
\eqno(2.3)$$
This addition to the action has the form of a magnetic coupling to an
external gauge field with potential $A_\mu = \A \omega_\mu +\B a_\mu$,
but in two dimensional spacetime it is possible to relate this gauge
field to the gravitational metric as in (1.4) and (2.3). This has the
consequence
that the point particle interacts with a gauge field whose curvature two-form
is everywhere proportional to the spacetime area two-form.

The effect of this additional matter-gravity coupling is to modify the usual
geodesic equation of motion arising from $I_m$ to contain a geometrical
force term and to read
$$
{d\over d\tau} {1\over N} \dot x^\mu + {1\over N} \dot x^\alpha \Gamma^
\mu_{\alpha \beta} \dot x^\beta +\left( \s{1\over 2} \A R + \B \right)
g^{\mu
\alpha}
\sqrt{-g} ~\epsilon_{\alpha \beta} \dot x^\beta =0
\eqno(2.4)$$
where $N\equiv {1\over m} \sqrt{\dot x^\alpha g_{\alpha \beta} \dot
x^\beta}$.

To understand the symmetries of the total action
$$
I \equiv I_m + I_{\rm coupling},
\eqno(2.5)$$
consider the variation of the total Lagrangian induced by a change $\delta
x^\mu$ in
the spacetime coordinates :
$$
\delta L = - {m\over2} {\dot x^\mu \dot x^\nu \over \sqrt{\dot
x^\alpha
  g_{\alpha\beta} \dot x^\beta}} \L_{\delta x} g_{\mu\nu}
- {d\over d\tau} \Bigl( (\A \omega_\mu + \B a_\mu) \delta x^\mu
\Bigr)
+ ( \s{1\over2} \A R + \B ) \sqrt{-g} \epsilon_{\mu\nu} \delta x^\mu
 \dot x^\nu
\eqno(2.6)$$
Here $\L_{\delta x}$ is the Lie derivative along $\delta x^\mu$. If
$\delta x^\mu$ is a Killing vector of the metric $g_{\mu \nu}$ ({\it
i.e.\/},
$\L_{\delta x} g_{\mu\nu}=0$ ), one can also
verify that $\sqrt{-g} \epsilon_{\mu\nu} \delta x^\mu =
\partial_\nu X$
for some function $X$. If moreover the curvature is constant, then the last
term
in $\delta L$ can be written as a total derivative; hence the action is
invariant
and the Killing vectors generate corresponding symmetries.

Let us thus concentrate on the case of a maximally symmetric two
dimensional
spacetime,
which has constant curvature $R = \Lambda$ and three Killing vectors
$$\eqalignno{
\xi^\mu_{(J)} &= ( -x^1, -x^0) \cr
\xi^\mu_{(0)}  &= \left( 1 + \s{\Lambda\over8}(x^0)^2 +
 \s{\Lambda\over8}(x^1)^2 , \s{\Lambda\over4} x^0 x^1 \right) &(2.7)\cr
\xi^\mu_{(1)}  &= \left( - \s{\Lambda\over4} x^0 x^1 ,
 1 - \s{\Lambda\over8}(x^0)^2 - \s{\Lambda\over8}(x^1)^2 \right)
\cr}
$$
Each Killing vector generates a symmetry of the action and is associated
with a conserved Noether ``current'' :
$$
Q = {\partial L\over \partial\dot x^\mu}
\delta x^\mu + ( \A \omega_\mu + \B a_\mu ) \delta x^\mu - ( \B +
\s{1\over2}\A \Lambda ) X
\eqno(2.8)$$
For the above Killing vectors, the conserved currents are
$$\eqalignno{
\delta x^\mu = - \xi^\mu_{(J)}: &\qquad
J = {- 1\over 1 + {\Lambda\over8} x^2} \left( {m\over\sqrt{\dot
x^2}}
\epsilon_{ab} x^a \dot x^b + \s{\BB\over2} x^2 \right) \cr
\delta x^\mu = - \xi^\mu_{(a)}: &\qquad
P_a = {m\over\sqrt{\dot x^2}} \dot x_a + \epsilon_{ab} x^b
(\BB + \s{\Lambda\over4} J) &(2.9)\cr}
$$
where $\BB \equiv \B + {1\over2} \A \Lambda$ and all indices are lowered
and raised with the flat metric $h_{ab} = \diag (1,-1)$.

Using the canonical symplectic structure $[{\delta L\over\delta\dot x^a},x^b]
=
\delta^b_a$, one computes the algebra of these currents to be
$$\eqalignno{
[P_a, J] &= \epsilon_a{}^b P_b \cr
[P_a, P_b] &= \epsilon_{ab} ( \s{\Lambda\over2} J + \BB ) &(2.10a)\cr}
$$
In the flat case, $\Lambda = 0$, one recognizes the extended Poincar\'e
algebra studied in~[4], and for $\Lambda \neq 0$ with $\BB=0$ (i.e. {\it
without}
the additional matter-gravity coupling) this is just the de Sitter algebra
studied
in~[6] in relation to two dimensional gravity.
We introduce a central element $I$, which is represented in this symplectic
representation by 1
$$
[P_a, P_b] = \epsilon_{ab} ( \s{\Lambda\over2} J + \BB I) \eqno (2.10b)
$$
and we shall refer to this symmetry
algebra as the ``extended de Sitter algebra". The above discussion
shows
that this algebra provides a direct bridge between the pure de Sitter and
extended Poincar\'e models. One can, of course, also redefine
the generator $J$ by adding a constant central element
$$
\JJ \equiv J + {2\BB\over \Lambda}I
\eqno(2.11)$$
in such a way that the resulting algebra is the conventional de Sitter one.
While this is often convenient for computations, the extended form (2.10) is
most useful for considering the contraction to the extended Poincar\'e
algebra.

Note also that these generators are related in the following way
$$
P_a h^{ab} P_b - \s{\Lambda\over2} J^2 - 2 \BB J I = m^2
\eqno(2.12)$$
The LHS of this relation is the quadratic Casimir of the extended de Sitter
algebra.

We shall see later that it is possible to construct a gauge formulation of
 this
gravitational coupling, with gauge algebra being precisely this extended de
Sitter algebra.
\vfill
\eject
\noindent {\bf II.2 First Order or Hamiltonian Formalism}
\medskip
It is instructive to formulate the above model in a first order formalism.
 This
will prove useful when considering the gauge formulation of the theory, and
is
also desirable for the eventual quantization of this theory.

The action (2.5) may be represented as
$$
I =  \int d\tau \left\{ (p_a e^a_\mu - \A \omega_\mu
- \B a_\mu) \dot x^\mu + {1\over2} N (p_a h^{ab} p_b - m^2) \right\}
\eqno(2.13)$$
Upon variation of the fields one obtains $p_a = - e^a_\mu \dot x^\mu /N$
and
$N=\sqrt{\dot x^\mu g_{\mu \nu} \dot x^\nu}/m$, in which case the
Lagrangian
form presented above is regained. Alternatively, one may write
$$
I = \int d\tau \left\{ \pi_\mu \dot x^\mu + {1\over2} N \Bigl( (\pi_\mu
+ \A
\omega_\mu + \B a_\mu) g^{\mu\nu} (\pi_\nu + \A \omega_\nu + \B
a_\nu) - m^2
\Bigr) \right\}
\eqno(2.14)$$
where $\pi_\mu$ is the momentum
$$\eqalignno{
\pi_\mu &= {\partial L \over \partial \dot x^\mu} \cr
&=p_a e^a_\mu - \A \omega_\mu - \B a_\mu &(2.15)\cr}
$$
This last form makes explicit the Hamiltonian of the system
$$
H = - {1\over2} N \Bigl( (\pi_\mu + \A \omega_\mu + \B a_\mu)
g^{\mu\nu}
 (\pi_\nu + \A \omega_\nu + \B a_\nu) - m^2 \Bigr)
\eqno(2.16)$$
In the first order formulation, a change $\delta x^\mu$ in the spacetime
coordinates is induced by the symplectic form $[\pi_\mu, x^\nu] =
\delta^\nu_\mu$ and the generator ${\cal G} \equiv \pi_\mu \delta x^\mu -
Y$,
where $Y(x)$ can be an arbitrary function of $x$ :
$$
\delta x^\mu = [\pi_\nu \delta x^\nu - Y, x^\mu]
\eqno(2.17)$$
The resulting change in the Hamiltonian is then
$$\eqalignno{
\delta H =& [\pi_\mu \delta x^\mu - Y, H] \cr
=& - N \Bigl( (\s{1\over2} \A R + \B) \sqrt{-g} \epsilon_{\mu\rho}
\delta
 x^\rho + \partial_\mu \left\{ (\A \omega_\rho + \B a_\rho) \delta x^\rho + Y
 \right\}
 \Bigr) g^{\mu\nu} \Bigl( \pi_\nu + \A \omega_\nu + \B a_\nu \Bigr) \cr
&- {1\over2} N \L_{\delta x} g^{\mu\nu} (\pi_\mu + \A \omega_\mu + \B a_\mu)
 (\pi_\nu + \A \omega_\nu + \B a_\nu) &(2.18)\cr}
$$
For constant curvature and $\delta x^\mu$ a Killing vector, the
variation of the Hamiltonian will vanish (thereby making $\G$ a symmetry
generator) provided one chooses
$$
Y = (\B + \s{1\over2} \A R) X - (\A \omega_\mu + \B a_\mu) \delta
x^\mu
\eqno(2.19)$$
where $X$ is defined as before by $\sqrt{-g} \epsilon_{\mu \nu}
\delta x^\mu =\partial_\nu X$.

If one applies this to the constant curvature case and the three
Killing vectors described earlier, one finds the generators
$$\eqalignno{
J &= - \pi_a \epsilon^a{}_b x^b \cr
P_a &= - (1 + \s{\Lambda\over8} x^2 ) \pi_a + {1\over2} (\BB +
\s{\Lambda\over2} J) \epsilon_{ab} x^b &(2.20)\cr}
$$
It is straightforward to verify that these generators satisfy the
same algebra (2.10) as that obtained above by Noether's theorem.
Moreover, the
Hamiltonian (2.16) can be expressed as a quadratic form in these generators
$$
H = - {1\over2} N \Bigl( P_a h^{ab} P_b - \s{\Lambda\over2} J^2 - 2\BB
J - m^2 \Bigr)
\eqno(2.21)$$
The expression in parentheses is simply the quadratic Casimir
of the extended de Sitter algebra, confirming indeed that the Hamiltonian is
invariant under transformations generated by this algebra.

For later use, we record the form of the conserved generators (2.20) in light
cone form:
$$\eqalignno{
P_+ &= \s{1\over\sqrt{2}} ( P_0 + P_1 )
= - \s{\Lambda\over4} (x^-)^2 \pi_- - \pi_+ + \s{\BB\over2} x^- \cr
P_- &= \s{1\over\sqrt{2}} ( P_0 - P_1 )
= - \s{\Lambda\over4} (x^+)^2 \pi_+ - \pi_- - \s{\BB\over 2}x^+ &(2.22)\cr
J &= x^+ \pi_+ - x^- \pi_- \cr}
$$
These generators satisfy the extended de Sitter algebra
$$\eqalignno{
[P_+,P_-] &= \s{\Lambda\over2} J + \BB I \cr
[P_\pm ,J]&=\pm P_{\pm} &(2.23)\cr}
$$

\bigskip
\noindent{\bf III. MATTER-GRAVITY COUPLING}
\bigskip
\noindent{\bf III.1 Equations of Motion}
\medskip
The equations of motion can be computed equivalently in both the
Lagrangian
and Hamiltonian formalisms. Let us consider the first one. The geodesic
equation
obtained by varying the action is
$$
{d\over d\tau} \left[ {m\dot x^\mu\over\sqrt{\dot x^\alpha g_{\alpha\beta}
\dot x^\beta}} \right] + {m\over\sqrt{\dot x^\alpha g_{\alpha\beta}
\dot x^\beta}}
 \Gamma^\mu_{\nu\rho} \dot x^\nu \dot x^\rho
+ (\s{1\over2} \A R + \B ) g^{\mu\nu} \sqrt{-g} \epsilon_{\nu\rho} \dot
x^\rho = 0
\eqno(3.1)$$
In the maximally symmetric case, with $R=\Lambda$ a constant, the
symmetry
generators (2.9) are first integrals of motion. Due to the algebraic relation
they satisfy, it is sufficient to consider two of them, namely
$$
P_a = {m\over\sqrt{\dot x^2}} \dot x_a + (\BB +
\s{\Lambda\over4} J) \epsilon_{ab} x^b \equiv (\BB + \s{\Lambda\over4} J)
\epsilon_{ab} \xx^b = {\rm constant}
\eqno(3.2)$$
where $\xx^a$ are constants of integration. This leads to classical
trajectories
of hyperbolic form :
$$
(\BB + \s{\Lambda\over4} J)^2 (x - \xx)^2  + m^2 = 0
\eqno(3.3)$$
Note that $J$ may be expressed in terms of the $\xx^a$ using the above Casimir
relation (2.12). We also remark that in the limit of flat spacetime,
$\Lambda \to 0$, the trajectory remains hyperbolic, as found in the
extended Poincar\'e case~[4].
\bigskip
\bigskip
\noindent{\bf III.2 Gravity Sector}
\medskip
We propose an action for gravity which enforces the condition of a
maximally symmetric spacetime and provides the needed one-form $a_\mu$
whose curvature two-form is everywhere proportional to the spacetime
area two-form. This was first considered in Ref.~12 and may be achieved using
two Lagrange multipliers $\eta,\lambda$:
$$
I_{\rm grav} = {1\over4\pi G} \int d^2x \sqrt{-g} \left[ \eta (R - \Lambda)
+ \lambda ( \s{1\over\sqrt{-g}} \epsilon^{\mu\nu} \partial_\mu a_\nu - 1)
\right]
\eqno(3.4)$$
The equations of motion obtained by varying with respect to the fields $\eta,
\lambda, g_{\mu \nu}, a_\mu$ are then
$$\eqalignno{
 R& = \Lambda \cr
 \s{1\over\sqrt{-g}} \epsilon^{\mu\nu} \partial_\mu a_\nu &= 1 \cr
(\nabla_\mu \partial_\nu - g_{\mu\nu} \nabla_\rho \partial^\rho) \eta &=
 \s{1\over2} \Lambda g_{\mu\nu} \eta + \s{1\over2} \lambda g_{\mu\nu}
&(3.5)\cr
 \partial_\mu \lambda&= 0 \cr}
$$
where $\nabla_\mu$ is the metric covariant derivative.

{}From the first two equations, we recover the geometric quantities
$$\eqalignno{
g_{\mu\nu}&= {h_{\mu\nu}\over(1 + {\Lambda\over8} x^2)^2} \cr
e^a_\mu&= {\delta^a_\mu\over1 + {\Lambda\over8} x^2} &(3.6)\cr
\omega_\mu = \s{\Lambda\over2} a_\mu&= \s{\Lambda\over4}
{\epsilon_{\mu\nu}
x^\nu\over1 + {\Lambda\over8} x^2} \cr}
$$
These agree (as they must) with the constant curvature solutions (1.5) and
(1.6) found before from the Liouville solution (1.9).
Notice that according to the last equation the field $a_\mu$ appears to be
redundant. This is true unless one wants to perform a $\Lambda \to 0$ limit;
besides providing an equation for $\lambda$, it is in that case necessary
for the gauge formulation.

The two last equations determine the Lagrange multiplier fields
$$\eqalignno{
\lambda&= \lambda_0 = \hbox{constant} \cr
\eta&= { {1\over2}\epsilon_{a b} x^a \varphi^b + (\varphi_2 +
{2\lambda_0\over\Lambda})
(1 - {\Lambda\over8} x^2) - {2\lambda_0\over\Lambda} \over 1 +
{\Lambda\over8} x^2} &(3.7)\cr}
$$
where $\lambda_0$, $\varphi^a$, $\varphi_2$ are integration constants.
Once again, in the flat space limit, $\Lambda \to 0$, the solutions of
the extended Poincar\'e theory are recovered.

For the sake of later comparison with the gauge formulation of this
gravitational theory, it is helpful to rewrite this gravity action (3.4)
in terms of the {\it Zweibein} and spin connection fields. To do so, we
recall the relation (1.4) between the curvature and the spin connection,
and we include a Lagrange multiplier to enforce the relation (1.2)
between the {\it Zweibein} and spin connection. Then the gravity
action becomes :
$$\eqalignno{
I_{\rm grav}={1\over4\pi G} \int d^2x ~\epsilon^{\mu \nu} \Bigl\{ 2\eta
(\partial_\mu
\omega_\nu + {\Lambda\over 4} e^a_\mu e^b_\nu \epsilon_{a b})&+
\lambda
(\partial_\mu
a_\nu + {1\over 2} e^a_\mu e^b_\nu \epsilon_{a b})\cr
& +\eta_a (\partial_\mu e^a_\nu + \epsilon^a{}_b \omega_\mu e^b_\nu)
\Bigr\}&(3.8)\cr}
$$
The equations of motion easily determine $e^a_\mu$, $\omega_\mu$ and
$a_\mu$ as above, and the equations of motion for the Lagrange
multiplier fields may be expressed most succinctly in light-cone notation:
$$\eqalignno{
\partial_\pm \lambda&=0\cr
\partial_\pm \eta &= \pm {{1\over 2}\eta_\pm \over {1+{\Lambda\over
8}x^2}}\cr
\partial_\pm \eta_\pm &= {-{\Lambda\over 4} x^{\mp} \eta_\pm \over
{1+{\Lambda\over 8}x^2}} &(3.9)\cr
\partial_\pm \eta_\mp &= {{\pm\left(\Lambda \eta + \lambda\right) +
{\Lambda\over 2}x^\mp \eta_\mp} \over {1+{\Lambda\over
8}x^2}}\cr}
$$
Integrating these equations yields the solutions
$$\eqalignno{
\lambda&=\lambda_0\cr
\eta&={1\over 2} \left( {x^+ \varphi^- - x^- \varphi^+ -
{2\alpha\over\Lambda}
\left(1-{\Lambda\over 8}x^2\right) \over {1+{\Lambda\over 8}x^2}} -
{2 \lambda_0 \over \Lambda} \right) &(3.10)\cr
\eta_\pm &= {\varphi^{\mp} + {\Lambda\over 4} \left(x^{\mp}\right)^2
\varphi^{\pm} \pm \alpha x^{\mp} \over {1+{\Lambda\over
8}x^2}} \cr}
$$
where $\lambda_0$, $\alpha$ and $\varphi^{\pm}$ are constants of
integration. Note that in order to compare with (3.7) and to obtain a finite
solution in the $\Lambda\to 0$ limit ({\it viz.} the extended Poincar\'e
theory~[4]) we can write the arbitrary constant $\alpha$ as
$$
\alpha=-\lambda_0 -\Lambda\varphi_2
\eqno(3.11)$$
where $\varphi_2$ is the integration constant in (3.7).

\vfill
\eject
\noindent{\bf IV. GAUGE FORMULATION OF GRAVITATIONAL THEORY}
\bigskip
\noindent{\bf IV.1 Invariant Inner Product for Extended de Sitter Algebra}
\medskip
As a precursor to the introduction of a gauge formulation for this
gravitational theory, we consider the extended de Sitter algebra
in more detail. Representing the four generators $P_a,J,I$ as $Q_A$, we can
express the commutation relations (2.10) as
$$
[Q_A,Q_B]=f_{AB}{}^{C}Q_C
\eqno(4.1)$$
and define an invariant inner product
$$
\bra Q_A|Q_B \ket = h_{A B}
\eqno(4.2)$$
The term ``invariant" means invariant under conjugation,
$$
\bra U^{-1}Q_A U|U^{-1}Q_B U\ket = \bra Q_A|Q_B \ket
\eqno(4.3)$$
This leads to the following form for $h_{A B}$:
$$
h_{A B} =\left(\matrix{ 1 & 0  & 0 & 0 \cr
                        0 & -1 & 0 & 0\cr
                        0 & 0  & c & - {1\over\BB} (1 + {c\Lambda\over2}) \cr
                        0 & 0  & - {1\over\BB} (1 + {c\Lambda\over2}) &
  {\Lambda\over2\BB^2} (1 + {c\Lambda\over2}) \cr}\right)
\eqno(4.4)$$
The inverse matrix $h^{A B}$ ($c \neq - {2\over\Lambda}$)
$$
h^{A B}=\left(\matrix{ 1 & 0  & 0 & 0 \cr
 		       0 & -1 & 0 & 0 \cr
		       0 & 0  & - {\Lambda\over2} & -\BB \cr
                       0 & 0  & -\BB & - {c\BB^2\over{1+c\Lambda/2}}
\cr}\right)
\eqno(4.5)$$
is used to define the quadratic Casimir operator
$$
C\equiv Q_A h^{A B} Q_B
\eqno(4.6)$$
which commutes with all the generators $Q_A$.

Note that if we choose the arbitrary parameter $c$ appearing in $h^{A B}$ and
$h_{A B}$ to be $c = \left({m\over\BB}\right)^2 / \left( 1 - {\Lambda\over2}
\left({m\over\BB}\right)^2 \right)$, the Casimir operator (4.6) will be
proportional to the Hamiltonian (2.21). However, it is perfectly consistent
(and more convenient) to set $c$ to zero.
This is achieved by shifting $J$ with an appropriate multiple of $I$,
and we make this choice
henceforth. Furthermore, notice that by setting $\Lambda=0$ the
extended de Sitter algebra reduces to the extended Poincar\'e algebra,
and the invariant
inner product reduces accordingly. Setting $\BB=0$, we get back to
the pure de Sitter case~[6], in which case the fourth generator $I$ decouples
and only a $3 \times 3$ sub-block of $h^{A B}$ is invertible, as expected.

\bigskip
\noindent{\bf IV.2 Extended de Sitter Gravity Action}
\medskip
We define a gauge field $A$ with values in the extended de Sitter
algebra (2.10),
$$
A = e^a P_a + \omega J + \BB a I
\eqno(4.7)$$
The corresponding field strength is
$$\eqalignno{
F &= dA + A^2\cr
&= (de^a + \epsilon^a{}_b \omega e^b) P_a +
 (d\omega + \s{\Lambda\over4} e^a \epsilon_{ab} e^b) J + \BB
 (da + \s{1\over2} e^a \epsilon_{ab} e^b) I &(4.8)\cr}
$$
Using the above invariant inner product we define the pure gravity
Lagrangian to be
$$\eqalignno{
L_{\rm grav}&= {1\over 4\pi G} \bra \eta | F\ket
= {1\over 4\pi G}\eta_A F^A \cr
&= {1\over 4\pi G} \{\eta_a (de^a + \epsilon^a{}_b \omega e^b)
 + \eta_2 (d\omega + \s{\Lambda\over4} e^a \epsilon_{ab} e^b)
 + \BB \eta_3 (da + \s{1\over2} e^a \epsilon_{ab} e^b)\} &(4.9)\cr}
$$
Then the equations of motion obtained by varying with respect to the
field $\eta$ may be expressed as
$$
F=0
\eqno(4.10)$$
Indeed, in a certain sense, the {\it key} to this type of gauge
formulation of gravity is the fact that under an infinitesimal
diffeomorphism $\delta x^\mu$ the gauge field changes according
to the Lie derivative formula
$$
\delta A_\mu = \delta x^\nu \partial_\nu A_\mu + (\partial_\mu \delta
x^\nu)
A_\nu
\eqno(4.11)$$
which may be rewritten as~[13]
$$
\delta A_\mu = \delta x^\nu F_{\nu \mu} + D_\mu (\delta x^\nu A_\nu)
\eqno(4.12)$$
so that ``on shell", where $F=0$, the effect of a diffeomorphism is
that of a gauge transformation ( with field dependent transformation
parameter ).

The $P_a$ projection of $F=0$ ,
$$
de^a + \epsilon^a{}_b \omega e^b =0
\eqno(4.13)$$
states that the torsion vanishes and defines the spin connection in
terms of the {\it Zweibein}. The $J$ projection of $F=0$
$$
d\omega = - \s{\Lambda\over 4} e^a \epsilon_{ab} e^b
\eqno(4.14)$$
implies [see (1.4)] that the scalar curvature is constant:
$$
R=\Lambda
\eqno(4.15)$$
Finally, the $I$ projection of $F=0$ implies that $a$ is proportional
to the spin connection, up to an arbitrary gauge transformation,
$$
a = \s{2\over \Lambda}\omega + d\theta
\eqno(4.16)$$
Using this last result we see that, ``on shell'', the gauge field $A$
may be re-expressed as
$$\eqalignno{
A&=e^a P_a +\omega (J + \s{2\BB\over \Lambda}I) +d\theta I\cr
&=e^a P_a +\omega \JJ +d\theta I&(4.17)\cr}
$$
where replacing $J$ by the {\it single} generator $\JJ \equiv J + {2\BB\over
\Lambda}I$ reduces the extended de Sitter algebra (2.10) to the
standard de Sitter algebra. The corresponding field strength is
$$
F=(de^a + \epsilon^a{}_b \omega e^b) P_a +
(d\omega + \s{\Lambda\over 4} e^a \epsilon_{ab} e^b) \JJ
\eqno(4.18)$$
\bigskip
\noindent{\bf IV.3 Canonical Structure of the Gravity Sector}
\medskip
To study the canonical structure of the gauge invariant gravity
Lagrangian (4.9), we first write the action in Hamiltonian form
$$\eqalignno{
I_{\rm grav} &= {1\over4\pi G} \int d^2x \epsilon^{\mu\nu} \eta_A
F^A_{\mu\nu}
\cr
&= {1\over4\pi G} \int dt\,dx ( \eta_A \partial_0 A^A_1 - A^A_0 \G_A )
 -  {1\over4\pi G} \int dt\,dx \partial_1 ( \eta_A A^A_0 ) &(4.19)\cr}
$$
where the constraints $\G_A$ are just the spatial covariant derivative of
$\eta_A$
$$
\G_A = -\left(\partial_1 \eta_A + f_{ABC} A^B_1 \eta^C \right)
\eqno(4.20)
$$
Here $f_{AB}{}^C = f_{ABD} h^{DC}$ are the structure constants of the algebra.
Then using the symplectic structure deduced from the action
$$
[\eta_A (x),A^B_1(y)] = 4 \pi G \, \delta^B_A \delta(x-y)
\eqno(4.21)$$
it is straightforward to check that the constraint algebra satisfies the
algebra
$$
[\G_A(x),\G_B(y)] = f_{AB}{}^C \G_C(x) \delta(x-y)
\eqno(4.22)$$
as is usual in gauge theories.

\bigskip
\noindent{\bf IV.4 Gauge Solution of The Equations of Motion}
\bigskip
In the gauge theoretical language it is particularly easy to solve
the equations of motion. This is because the general
solution for the gravitational fields, see eq. (4.10), is
$$\eqalignno{
A&= {\rm pure~gauge}\cr
&=U^{-1}dU&(4.23)\cr}
$$
for some group element $U$. A simple computation shows that by taking
$$
U = e^{x^+ P_+} e^{-\log (1+{\Lambda\over 8}x^2) \JJ} e^{x^- P_-}
\eqno(4.24)$$
one finds that
$$\eqalignno{
A&=U^{-1}dU \cr
&= e^a P_a + \omega J + \BB a I&(4.25)\cr}
$$
where $e^a$, $\omega$ and $a$ coincide
with the {\it Zweibein}, spin connection and gauge field $a$ given in (3.6).

The equations of motion for the Lagrange multiplier fields $\eta$ may
be succinctly expressed as
$$
D\eta =0
\eqno(4.26)$$
where $D$ refers to the covariant derivative with respect to the
gauge field $A$ in (4.17). Thus, with $A=U^{-1}dU$, the solution
for $\eta$ is just
$$
\eta = U^{-1} \eta_{(0)} U
\eqno(4.27)$$
where $\eta_{(0)}$ is some constant element of the extended de
Sitter algebra. Indeed, expanding an arbitrary constant algebra element as
$$
\eta_{(0)} = \varphi^+ P_+ +\varphi^- P_- + \alpha \bar{J} + \beta I
\eqno(4.28)$$
( note that for algebraic convenience we use $\JJ \equiv J +
{2\BB\over\Lambda}I$ instead of $J$) one finds that
$$\eqalignno{
\eta =& U^{-1} \eta_{(0)} U\cr
=& \left({{\varphi^+ + {\Lambda\over 4}(x^+)^2 \varphi^- -
\alpha x^+}\over {1+{\Lambda\over 8}x^2}}\right) P_+
+\left({{\varphi^- + {\Lambda\over 4}(x^-)^2 \varphi^+ + \alpha x^-}\over
{1+{\Lambda\over 8}x^2}}\right)P_-\cr
&-{\Lambda\over 2}\left({{x^+ \varphi^- -x^- \varphi^+ -
{2\alpha\over\Lambda}\left(1-{\Lambda\over 8}x^2\right)}\over {1+{\Lambda\over
8}x^2}}
\right)\bar{J} +\beta I&(4.29)\cr}
$$
If we write the arbitrary integration constant $\beta$ as
$-2\lambda_0 /\Lambda$, then it is straightforward to use
$\eta_A =h_{A B}\eta^B$ to show that this solution for $\eta$
coincides with that obtained before in (3.10) by integrating
directly the equations of motion.

\bigskip
\noindent{\bf V. GAUGE COVARIANT MATTER-GRAVITY COUPLING}
\bigskip
\noindent{\bf V.1 de Sitter Gauge Invariant Action}
\bigskip
To discuss the gauge covariant form of the matter-gravity coupling
it is most convenient to work with the first order formalism of
Section II.2. To facilitate comparison with earlier papers~[5], we redefine
the fields $p_a$ in the action (2.13) as $-\epsilon_a{}^b p_b$, so that the
first order action reads
$$
I_{\rm matter}=-\int d\tau \left\{\left(p_a\epsilon^a{}_b e^b_\mu
+\A\omega_\mu +
\B a_\mu\right)\dot{x}^\mu + {1\over2}N\left(p^2+m^2\right)\right\}
\eqno(5.1)$$
This is Lorentz invariant, but not de Sitter invariant ( due to
the lack of translational invariance of the {\it Zweibein} $e^a_\mu$ ).
It is, however, possible to write this action in a de Sitter
invariant form using the gauge formulation. We shall use a variant of
a technique due to Grignani and Nardelli~[14] in which the factor
$-\epsilon^a{}_b e^b_\mu \dot x^\mu$ is replaced by the gauge
covariant derivative (with respect to $\tau$) of a set of auxiliary
variables $q^a$. The idea is to introduce these extra degrees of
freedom in a gauge covariant manner and so that the action (5.1) is
obtained by a particular gauge choice. A similar idea has been studied in the
context of four dimensional de Sitter gravity theories by Stelle and
West~[15].

Specifically, we define elements $q$ and $p$ of the extended de Sitter
algebra.
Thus $p$ and $q$ transform as
$$\eqalignno{
q&\to U^{-1} q U \cr
p&\to U^{-1} p U &(5.2)\cr}
$$
Furthermore, we choose $q$ and $p$ to be of the form
$$\eqalignno{
q&\equiv T^{-1} q_{(0)} T\cr
p&\equiv T^{-1} p_{(0)} T&(5.3)\cr}
$$
where $q_{(0)}$ is a {\bf constant} algebra element with components just in the
$J$ and $I$ directions
$$
q_{(0)} = J + \A I,
\eqno(5.4)$$
where $p_{(0)}$ has components just in the de Sitter translation directions
$$
p_{(0)} =p^a P_a,
\eqno(5.5)$$
and where $T$ is a generator of a de Sitter translation
$$
T=e^{-\xi^a \epsilon_a{}^b P_b}
\eqno(5.6)$$
We shall refer to the parameters $\xi^a$ which characterize $T$ as ``de Sitter
coordinates". The algebra elements $q$ and $p$ in Eq. (5.3) depend upon the
four degrees of freedom characterized by $\xi^a$ and $p^a$. These
relations,
(5.2) to (5.6), define a nonlinear action~[15] of the de Sitter group on the de
Sitter coordinates $\xi^a$. That is, under the transformation (5.2) we deduce
that $T$ transforms as
$$
T\to T^\prime \equiv e^{-\xi^{\prime a} \epsilon_a{}^b P_b}
\eqno(5.7)$$
where ${\xi^a}^\prime$ is a nonlinear function of the original de Sitter
coordinates $\xi^a$ and the parameters of the transformation group element.

Now consider the action
$$
I=\int d\tau \{ \bra p | D_{\tau}q \ket +\BB \bra q | A-
 T^{-1}\dot{T}\ket -{1\over 2}N(\bra p|p\ket +m^2)\}
\eqno(5.8)$$
where $A\equiv A_\mu \dot{x}^\mu$. Under a gauge transformation,
$$
A\to U^{-1} A U + U^{-1} \dot U
\eqno(5.9)$$
the action (5.8) is gauge invariant ( with surface terms taken to vanish ).
Moreover, if we transform to the
gauge in which
$$
q=q_{(0)}
\eqno(5.10)$$
this has the effect of removing the de Sitter coordinates. In this gauge,
$p=p_{(0)}$, and
$$\eqalignno{
A&\equiv A_{(0)}\cr
&=(e^a_\mu P_a + \omega_\mu J + \BB a_\mu I ) \dot{x}^\mu &(5.11)\cr}
$$
Then
$$\eqalignno{
\bra p | D_{\tau} q \ket & = \bra p_{(0)} | \dot{q}_{(0)}
+[A_{(0)},q_{(0)}]\ket\cr
&= - p_a \epsilon^a{}_b e^b_{\mu} \dot{x}^\mu &(5.12a)\cr}
$$
and
$$\eqalignno{
\BB \bra q|A- T^{-1}\dot T \ket &=\BB \bra q_{(0)} | A_{(0)} \ket \cr
&=-(\A \omega_\mu + \B a_\mu )\dot{x}^\mu &(5.12b)\cr}
$$
so that the action (5.8) reduces to the original action (5.1).

This is analogous~[5] to the Higgs phenomenon in vector gauge theories.
There, given the gauge noninvariant Lagrange density
$$
\L = -{1\over 4} F^{\mu \nu}F_{\mu \nu} +{1\over 2} \partial_\mu \rho
\partial^\mu \rho +{1\over 2} \rho^2 A_\mu A^\mu - V(\rho)
\eqno(5.13)$$
one may introduce an additional variable $\theta$ (the analogue of the de
Sitter
coordinates $\xi^a$), define $\varphi\equiv {1\over \sqrt{2}}\rho e^{i \theta}$
and then consider the gauge invariant Lagrange density
$$
\L'= -{1\over 4} F^{\mu \nu}F_{\mu \nu}
+(\partial_\mu-iA_\mu)\varphi^*(\partial^\mu +iA^\mu)\varphi -
V(2\sqrt{\varphi^*\varphi})
\eqno(5.14)$$
The original Lagrange density may be regained either by a choice of gauge
(set $\theta$ to zero and define $\varphi \equiv {1\over \sqrt{2}}\rho$), or
alternatively, in any gauge, by redefining variables in $\L'$ as $\varphi
={1\over\sqrt{2}}\rho e^{i \theta}$, $\tilde{A_\mu}=A_\mu +\partial_\mu
\theta$. In the example considered in this section, we only need to introduce
additional degrees of freedom $\xi^a$ corresponding to the de Sitter
translation
generators $P_a$ since the action (5.1) is already Lorentz invariant.

\vfill
\eject
\noindent{\bf V.2 Gauge Equations of Motion}
\bigskip
Including both the matter and gravity contributions, the total gauge
theoretical action is
$$
I_{\rm total}=\int d\tau \left\{\bra p|D_\tau q\ket +\BB \bra q|A-T^{-1}\dot T
\ket -{1\over2}
N( \bra p |p\ket +m^2 ) \right\} +{1\over 4\pi G}\int \bra \eta |F\ket
\eqno(5.15)$$
Since $q\equiv T^{-1}q_{(0)} T$ we can write the first term in the action as
$\bra [q,p]|A-T^{-1} \dot T \ket$ so that the action may be expressed as
$$
I_{\rm total}=\int \left\{\bra \BB q+[q,p]|A-T^{-1}\dot T \ket -{1\over 2}
N( \bra p |p\ket +m^2 )\right\} +{1\over 4\pi G}\int \bra \eta |F\ket
\eqno(5.16)$$
This second, equivalent, form of the action is useful for obtaining some of
 the
equations of motion. Variation with respect to the fields $\eta$, $N$, $p$,
$\xi^a$ and
$A_\mu$ lead to the equations of motion:
$$\eqalignno{
F&=0&(5.17 a)\cr
\bra p|p\ket +m^2&=0&(5.17 b)\cr
D_\tau q - Np&=0&(5.17 c)\cr
\BB D_\tau q - [ D_\tau p , q ] &= 0 & (5.17d) \cr
\epsilon^{\mu \nu} D_\nu \eta&=4\pi G \int d\tau
\Bigl(\BB q(\tau)+[q(\tau),p(\tau)]\Bigr) \dot x ^\mu \delta ^2
(x-x(\tau))&(5.17 e)\cr}
$$
Note that in the gauge $\xi^a = 0$ Eq.~(5.17$d$), obtained by varying with
respect to the de Sitter coordinates $\xi^a$, does not
lead to an independent equation of motion (see Ref.~[5]). This statement is
in fact true in any gauge.

The first three of these equations (5.17$a-c$) are easily solved, as before, to
give
$$\eqalignno{
A&=U^{-1}dU\cr
N&={1\over m}\sqrt{-\bra D_\tau q|D_\tau q\ket} &(5.18)\cr
p&={m D_\tau q \over \sqrt{-\bra D_\tau q|D_\tau q\ket}}\cr}
$$
In the special gauge (5.10), (5.11), $U$ is given by (4.24) and $D_\tau q=-
\epsilon^a{}_be^b_\mu \dot x ^\mu P_a$, so that
$$\eqalignno{
N&={1\over m}\sqrt{\dot x^\mu g_{\mu \nu} \dot x^\nu}\cr
p&= - {\epsilon^a{}_be^b_\mu \dot x ^\mu P_a\over N}&(5.19)\cr}
$$
as is seen directly from the action (5.1). Equation (5.17$e$) is a little more
subtle. Off the world line, where the delta function vanishes, we have
$D_\mu\eta=0$, whose solution is presented in (4.29). Then, in the
parametrization in
which $x^0(\tau)=\tau=t$, we may integrate across the world-line to obtain
the
solution
$$\eqalignno{
\eta &= U^{-1}(x) \Bigl(\eta_{(0)} + \Delta\eta_{(0)} \,
\varepsilon(x^1-x^1(t)) \Bigr) U(x) \cr
\Delta\eta_{(0)} &= 4\pi G U(x(t)) \Bigl( \BB q(t)+[q(t),p(t)]\Bigr)
U^{-1}(x(t)) & (5.20) \cr}
$$
where $\eta_{(0)}$ is the matter-free solution in (4.28) and
$\Delta\eta_{(0)}$
is a {\bf constant} depending on the parameters (3.2) of the particle
trajectory
(this is easily shown using the identity $D_\tau ( \BB q + [q,p] ) = 0$).
We can then eliminate
the inessential $\xi^a$ degrees of freedom by going to the the gauge
(5.10), (5.11)
 in
which case
$$
\BB q(t)+[q(t),p(t)] = - {e^a_\mu \dot x^\mu P_a\over N} + \BB ( J + \A I )
\eqno(5.21)$$
We see thus that the constants $\phi^a$, $\alpha$, and $\beta$ in (4.28),
(4.29) get merely shifted when one crosses the particle trajectory.

\bigskip
\noindent{\bf VI. CONCLUDING REMARKS}
\bigskip
We conclude by briefly summarizing our results. We have given a gauge
formulation for gravitational forces experienced by point particles in two
dimensional spacetime. For the maximally symmetric case of point particles
interacting with a constant (in general, nonzero) curvature gravitational
background there are three conserved generators (corresponding to the
three spacetime Killing vectors), which satisfy the de Sitter symmetry algebra.
When
we include the novel matter-gravity interaction in which the matter is
minimally coupled to a gauge field with gauge curvature correlated with the
spacetime metric curvature, the symmetry generators are modified and the
symmetry algebra acquires a central extension. Formally, this is analogous to
the idea of ``magnetic translation" symmetry, generalized to curved surfaces.
The resulting theory, based on this ``extended de Sitter algebra",
encompasses previous gauge theoretical models of two dimensional gravitational
theories. In the limit in which the new matter-gravity coupling is removed one
regains the pure de Sitter theory of Ref. [6]. And in the limit of vanishing
spacetime curvature, the extended de Sitter model contracts smoothly to the
extended Poincar\'e theory~[4,5].

In these theories, one effectively trades spacetime diffeomorphism
invariance for on shell gauge invariance.The significance of the gauge
formulation has been illustrated here by studying the symmetry generators and
the equations of motion. First, one sees that in a purely geometrical
formulation of this model (in terms of metric quantities), the extended de
Sitter algebra arises as a Noether symmetry algebra. Then one notices that it
is possible to define a gauge field with values in this same algebra in such a
way that a simple gauge action yields the original geometric theory. Being a
first order (in derivatives) action, the gauge theory equations of motion are
easier to solve and, moreover, the gauge structure provides a natural and
concise algebraic set of fields with which to analyze the model. This should
also play an important role in the quantization of this model, a question to be
addressed in future work.
In this study we have focused our attention on the point particle
interaction. The coupling of matter fields, like fermions, is also possible
and its gauge formulation can be carried over following the same ideas than
in Ref.~5.

Contact with the geometric formulation of dilaton gravity~[2] is achieved
by considering the ``stringy'' metric $\bar g_{\mu\nu} = e^a_\mu h_{ab}
e^b_\nu / \eta$. In the extended Poincar\'e case ($\Lambda = 0$), one
then recovers the well-known black-hole configuration, where the quadratic
expression (3.7) for $\eta$ describes the location of the horizon and the
``mass'' of the black-hole~[4]. The generalization of this connection to the
extended de Sitter case ($\Lambda\neq 0$) is an interesting question which also
deserves future attention.
\bigskip
\bigskip
\bigskip
\bigskip
\noindent{\bf Acknowledgements :}
\bigskip
We are grateful to Roman Jackiw for helpful discussions and comments. DC is
supported in part by the D.O.E. under Grant
DE-AC02-76ER03069 and by the Fonds du 450e de l'Universit\'e de
Lausanne, and GD is supported in part by the D.O.E. under Grant DE-FG02-
92ER40716.00.
\vfill
\eject
\centerline{\bf REFERENCES}
\bigskip
\item{[1]} C.~Teitelboim, {\it Phys. Lett. B} {\bf 126}, 41 (1983);
R.~Jackiw, in {\it Quantum Theory of Gravity}, S.~Christensen, ed.
(Adam Hilger, Bristol, 1984); R.~Jackiw, {\it Nucl. Phys. B} {\bf 252}, 343
(1985).
\medskip
\item{[2]} C.~Callan, S.~Giddings, J.~Harvey and A.~Strominger,
{\it Phys. Rev. D} {\bf 45}, 1005 (1992).
\medskip
\item{[3]} H.~Verlinde, in {\it Sixth Marcel Grossmann Meeting on
General Relativity}, M.~Sato, ed. (World Scientific, Singapore, 1992).
\medskip
\item{[4]} D.~Cangemi and R.~Jackiw, {\it Phys. Rev. Lett.} {\bf 69}, 233
 (1992); R.~Jackiw, {\it Theor. Math. Phys.} {\bf 9}, 404 (1992).
\medskip
\item{[5]} D.~Cangemi and R.~Jackiw, {\it Phys. Lett. B} {\bf 299}, 24
 (1993); D.~Cangemi and R.~Jackiw, {\it Ann. Phys. (NY)} {\bf 225}, 229
 (1993).
\medskip
\item{[6]} T.~Fukiyama and K.~Kamimura, {\it Phys. Lett. B} {\bf 160}, 259
(1985);
K.~Isler and C.~Trugenberger, {\it Phys. Rev. Lett.} {\bf 63}, 834 (1989);
A.~Chamseddine and D.~Wyler, {\it Phys. Lett. B} {\bf 228}, 75 (1989).
\medskip
\item{[7]} For a review, see : R. Mann, in {\it Proceedings of the Fourth
Canadian Conference on General Relativity and Relativistic
Astrophysics}, to be published.
\medskip
\item{[8]} J. Harvey and A. Strominger, Quantum Aspects of Black Holes
EFI-92-41, hep-th/9209055 (September 1992).
\medskip
\item{[9]} D.~Cangemi, {\it Phys. Lett. B} {\bf 297}, 261 (1992);
A. Ach\`ucarro, {\it Phys. Rev. Lett.} {\bf 70}, 1037 (1993);
G. Grignani and G. Nardelli, ``Poincar\'e Gauge Theories of Lineal Gravity",
Perugia University preprint DFUPG-57-1992 (August 1992); S. Kim, K. Soh and
J.~Yee, {\it Phys. Lett. B} {\bf 300}, 223 (1993).
\medskip
\item{[10]} A.~Comtet, {\it Ann. Phys. (NY)} {\bf 173}, 185 (1987);
G.~Dunne, {\it Ann. Phys. (NY)} {\bf 215}, 233 (1992).
\medskip
\item{[11]} D.~Hwang, S.~Kim, K.~Soh and J.~Yee, ``Particle Motion in a
Curved Cylindrical Geometry'', to appear in {\it Phys. Rev. D};
 R.~Iengo and D.~Li, ``Quantum Mechanics and Quantum Hall Effect
 on Riemann Surfaces", SISSA Preprint 100/93 (July 93).
\medskip
\item{[12]} S.~Kim, K.~Soh and J.~Yee, {\it Phys. Rev. D} {\bf 47}, 4433
(1993).
\medskip
\item{[13]} R.~Jackiw, {\it Phys. Rev. Lett.} {\bf 41}, 1635 (1978).
\medskip
\item{[14]} G.~Grignani and G.~Nardelli, in Ref. 11, and ``Canonical Analysis
of Poincar\'e Gauge Theories of 2D Gravity", Perugia University Preprint
DFUPG-76-1993.
\medskip
\item{[15]} K.~Stelle and P.~West, {\it Phys. Rev. D} {\bf 21}, 1466 (1980).
\medskip
\vfill
\bye